\documentclass{WileyMSP-template}
\begin{document}

\pagestyle{fancy}
\rhead{\includegraphics[width=2.5cm]{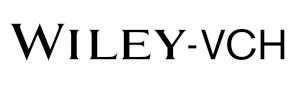}}

\title{Mobile Kink Solitons in a Van der Waals Charge-Density-Wave Layer}

\maketitle


\author{Jinwon Lee}
\author{Jae Whan Park}
\author{Gil-Young Cho}
\author{Han Woong Yeom*}



\begin{affiliations}
Dr. Jinwon Lee\\
Center for Artificial Low Dimensional Electronic Systems, Institute for Basic Science (IBS), Pohang 37673, Republic of Korea\\
Department of Physics, Pohang University of Science and Technology, Pohang 37673, Republic of Korea\\
Current Address: Leiden Institute of Physics, Leiden University, Leiden 2333 CA, The Netherlands\\
Email Address: jinwon.lee@physics.leidenuniv.nl
\medskip

Dr. Jae Whan Park\\
Center for Artificial Low Dimensional Electronic Systems, Institute for Basic Science (IBS), Pohang 37673, Republic of Korea\\
Email Address: absolute81@ibs.re.kr
\medskip

Prof. Dr. Gil-Young Cho\\
Center for Artificial Low Dimensional Electronic Systems, Institute for Basic Science (IBS), Pohang 37673, Republic of Korea\\
Department of Physics, Pohang University of Science and Technology, Pohang 37673, Republic of Korea\\
Email Address: gilyoungcho@postech.ac.kr
\medskip

Prof. Dr. Han Woong Yeom\\
Center for Artificial Low Dimensional Electronic Systems, Institute for Basic Science (IBS), Pohang 37673, Republic of Korea\\
Department of Physics, Pohang University of Science and Technology, Pohang 37673, Republic of Korea\\
Email Address: yeom@postech.ac.kr

\end{affiliations}



\begin{abstract}

Kinks, point-like geometrical defects along dislocations, domain walls, and DNA, are stable and mobile, as solutions of a sine-Gordon wave equation. While they are widely investigated for crystal deformations and domain wall motions, electronic properties of individual kinks have received little attention. In this work, electronically and topologically distinct kinks are discovered along electronic domain walls in a correlated van der Waals insulator of 1\textit{T}-TaS\textsubscript{2}. Mobile kinks and antikinks are identified as trapped by pinning defects and imaged in scanning tunneling microscopy. Their atomic structures and in-gap electronic states are unveiled, which are mapped approximately into Su-Schrieffer-Heeger solitons. The twelve-fold degeneracy of the domain walls in the present system guarantees an extraordinarily large number of distinct kinks and antikinks to emerge. Such large degeneracy together with the robust geometrical nature may be useful for handling multilevel information in van der Waals materials architectures.

\end{abstract}


\section{Introduction}

When a discrete symmetry is broken in an electronic system, degenerate ground states are generated.
They would coexist to have an inhomogeneous phase landscape, as laterally connected by various line defects of domain walls (DWs).
These line defects have provided intriguing physics such as topological edge modes in topological materials and interesting functionality in various ferroic materials for spintronic~\cite{Seidel2016,He2022} and electronic applications~\cite{Cheon2015,Yin2016,KimTH2017,Li2020,ParkJH2022}.
DWs in interacting electronic systems can have distinct localized states, which often govern physical properties and may lead to unprecedented functionality and novel devices~\cite{Catalan2012}. Outstanding examples are conductive DWs in multiferroic insulators~\cite{Seidel2009,Meier2012,Oh2015}, magnetic insulators~\cite{Yamaji2014,Ma2015}, Mott insulators~\cite{Kivelson1998}, and layered charge-density-wave (CDW) materials~\cite{Sipos2008,Joe2014,Yu2015,Li2016}. However, electronic states within DWs themselves have not been clearly identified and understood for most of the above systems.

Similar to crystal dislocations, these DWs can have higher order topological defects such as kinks.
For example, kinks in magnetic DWs has been widely studied for the motion of DWs~\cite{Parkin2008,Perez-Junquera2008,Buijnsters2014}.
DWs with well isolated one-dimensional (1D) electronic channels can also have kinks.
Since breaking translational symmetry in certain 1D electronic channels would lead to the formation of electronic solitons, kinks in electronic DWs may lead to the formation of electronic solitons.
However, the materials realization of an electronic kink soliton along a DW has been elusive.

An electronic soliton is described by the Su-Schrieffer-Heeger Hamiltonian, which features degenerate 1D ground states interfaced by soliton DWs~\cite{Su-Schrieffer-Heeger1979,Su-Schrieffer-Heeger1980}.
A soliton is endowed with a localized state within the band gap, which is protected by the topology given by the degeneracy of the broken symmetry, namely, \textit{Z}\textsubscript{2} symmetry for the dimer chain~\cite{Su-Schrieffer-Heeger1979,Su-Schrieffer-Heeger1980}.
Electronic solitons have been observed in highly-anisotropic quasi 1D crystals and organic polymers traditionally~\cite{Su-Schrieffer-Heeger1979,Su-Schrieffer-Heeger1980,Su-Schrieffer1981,Heeger1988,Hernangomez-Perez2020} and, more recently, in self-assembled atomic chains~\cite{Cheon2015,KimTH2017,ParkJH2022,KimTH2012,LeeGS2019}, artificial lattices~\cite{Huda2020}, and cold atoms~\cite{Atala2013,He2018,Chen2018}. 
For self-assembled atomic chains on crystal surfaces, individual solitons and their motions were identified by scanning tunneling microscopy ~\cite{Cheon2015,KimTH2017,ParkJH2022,KimTH2012,LeeGS2019}, which further 
suggested the possibility of utilizing individual solitons to carry, store, and manipulate information.
This type of informatics is topologically protected and, possibly, dissipation-less. However, the limitation of subtle and fragile material systems in artificial or self-assembled atomic chain systems is serious enough to prevent many types of measurements and applications other than ultra-high vacuum probe microscopy.  
It is noteworthy that the technology of using optical solitons for the efficient long-distance delivery of classical digital information has been well established in optical communication~\cite{Zysset1987,Dimitre2003}, while these solitons are not topologically protected.
Thus, a much more robust and easily accessible material is highly requested for topological soliton informatics.

In this work, we identify kink and antikink solitons in electronic DWs of a two-dimensional (2D) van der Waals material, which is relatively robust and integrable into various device architectures.
In addition to the materials merit, kinks in a 2D material have geometrically \textit{Z} topology to allow the accumulation of a large number of kinks in stark contrast to the \textit{Z}\textsubscript{2} solitons of 1D systems.
These properties make the present kink solitons an attractive platform to pursue the possibility of robust multilevel computations~\cite{Hatami-Hanza1997,Akhmediev2000,Seo2020}.

\section{Results and Discussion}
\subsection{DWs and mobile kinks in 1\textit{T}-TaS\textsubscript{2}}
\begin{figure}
  \includegraphics[width=17.8 cm]{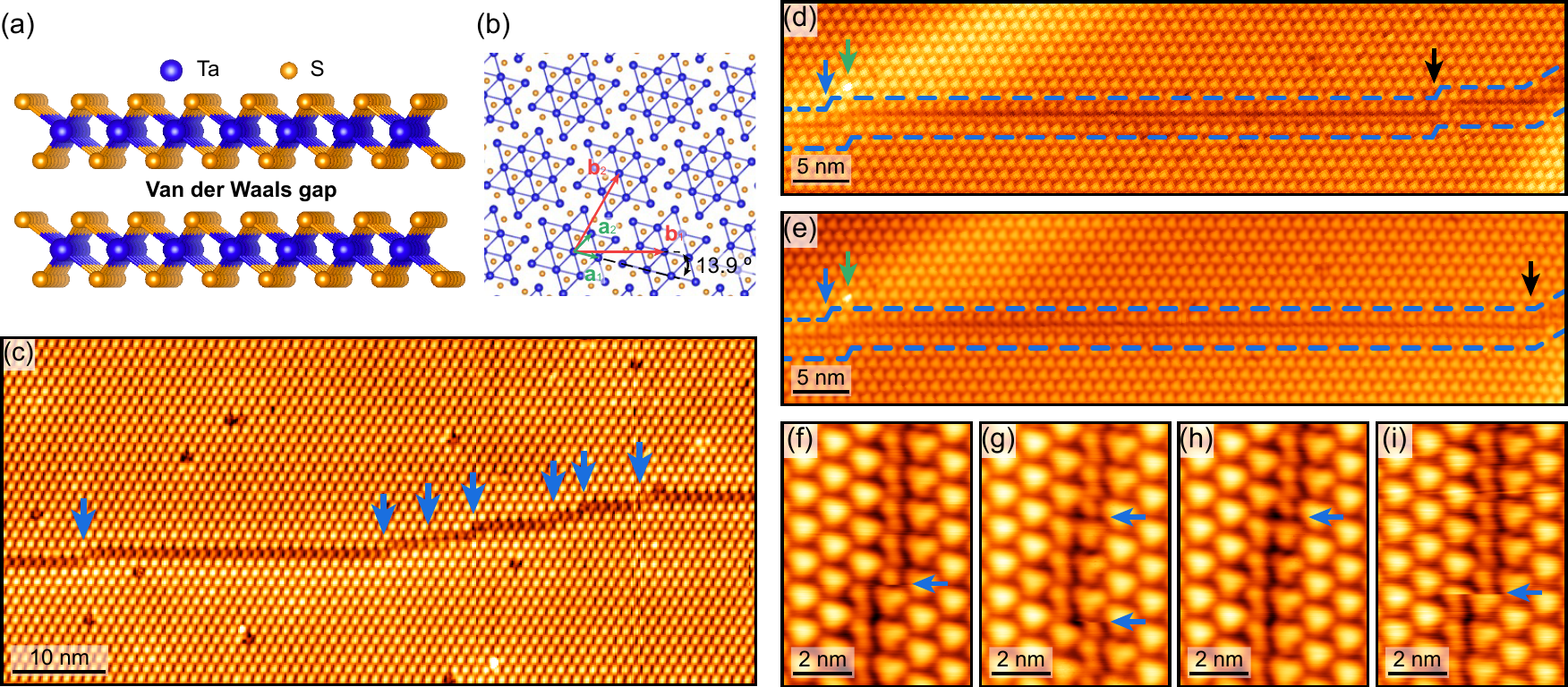}
  \caption{Mobile kinks along domain walls in 1\textit{T}-TaS\textsubscript{2}.
  (a)~Layered 1\textit{T}-TaS\textsubscript{2} crystal structure with van der Waals gap.
(b)~David-star CDW unitcell.
(c)~Kinks in a straight DW of 1\textit{T}-TaS\textsubscript{2}.
(d) and (e)~STM images of a kinked DW with a time interval of 1 hour.
One of two kinks in (d), marked with a black arrow, moves toward right and is trapped at the corner of the DW in (e).
(f)-(i)~Selected STM images from a sequence of scans at the same field of view (10 minutes for each image for about 17 hours), which show snap shots of mobile kinks.
}
  \label{mobile}
\end{figure}

Layered transition metal dichalcogenide 1\textit{T}-TaS\textsubscript{2} [Figure~\ref{mobile}(a)] provides 1D electronic channels formed along DWs.
It exhibits a series of symmetry-breaking phase transitions to insulating CDW phases at low temperature~\cite{Wilson1975} and various DWs related to the CDW phase were identified with well defined in-gap electronic states~\cite{Cho2017,ParkJH2021}.
In the ordered CDW phase, 13 Ta atoms are distorted to form a David-star-shaped unitcell in a $\sqrt{13}\times\sqrt{13}$ superstructure [Figure~\ref{mobile}(b)].
The origin of the insulating CDW phase can be either the on-site Coulomb repulsion~\cite{Cho2017,Fazekas1979,Pillo2000,Rossnagel2005,Sipos2008,Lahoud2014,Cho2015,Cho2016,Law2017,Ribak2017,Ligges2018,Park2019,LeeJW2020,Butler2020,LeeJW2021} or the layer dimerization depending on the stacking order~\cite{LeeJW2020,Butler2020,LeeJW2021,Ritschel2015,Ritschel2018,LeeSH2019,Stahl2020,Wang2020}, but the different origins of the gap are not important for the present discussion focusing on in-gap electronic states of DWs.  
The previous scanning tunneling microscopy and spectroscopy (STM and STS) studies identified various DWs between 13 degenerate CDW states~\cite{Cho2017,ParkJH2021,Cho2016,Fujii2018,Ma2016,Skolimowski2019}.
The STM images in Figure~\ref{mobile}(c)-\ref{mobile}(e) show one of the two major types of them, named DW4~\cite{ParkJH2021}.
As shown clearly here, some DWs can have quite a few (at least more than one in 100~nm) kinks (blue arrows), at which the straight DW is shifted by one CDW unitcell. 
Some kinks are mobile as detected in the successive scans in Figure~\ref{mobile}(d) and \ref{mobile}(e); it jumps abruptly from one to the other sites (black arrows).
Their mobility is well visualized in a sequence of scans with the same field of view over several hours, of which snap shots are illustrated in Figure~\ref{mobile}(f)-\ref{mobile}(i) (see also Supplemental Video~1).
The motion of the mobile kinks is so fast that we were not able to investigate the dynamic details of their motion due to a technical limitation of the relatively long timescale of the STM measurements.
Nevertheless, we could see an abrupt change in the topographic signal (an discontinuity of the DW) during the scan [Figure~\ref{mobile}(i)], which indicates a short timescale of the kink dynamics; the kink is much faster than the scanning tip whose typical speed is 10~nm/sec.
Note that we can rule out an abrupt change of our STM tip because it would have caused discontinuity in CDW unitcells while the unitcells are smoothly connected in the above observation.
The static kinks, on which we are focused in the rest of this article, can be associated with pinning defects nearby [the green arrows in Figure~\ref{mobile}(d) and \ref{mobile}(e)].
The fast diffusion of solitons and their trapping by defects have been well documented in the previous STM works~\cite{Cheon2015,KimTH2017,ParkJH2022,KimTH2012,LeeGS2019}.


\subsection{Atomic structures of a kink and an antikink}
\begin{figure}
  \includegraphics[width=16.8 cm]{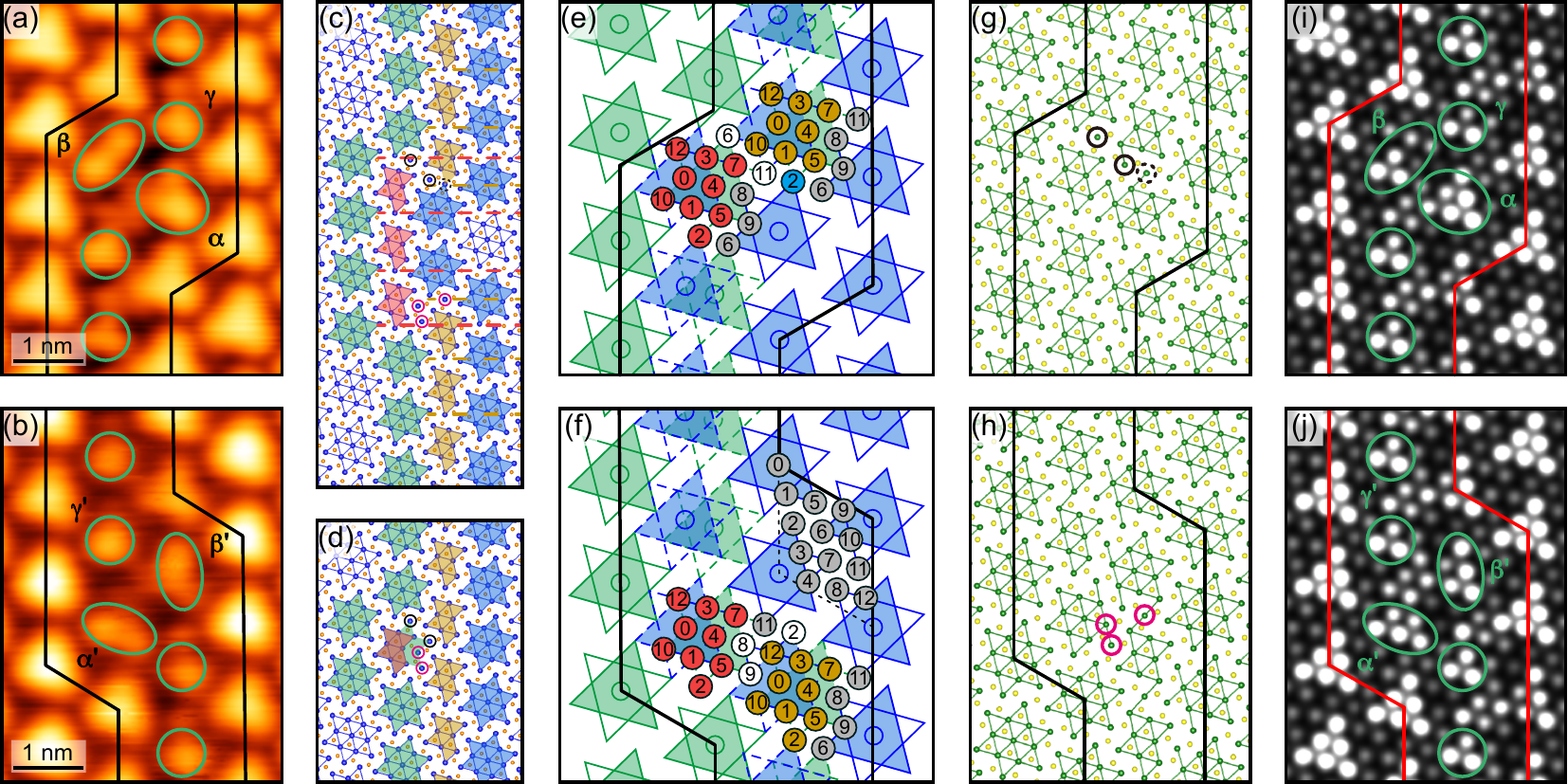}
  \caption{Atomic structures of a kink and an antikink along DW4.
(a) and (b)~STM images of a kink and an antikink along a DW4-type domain wall, respectively.
(c)~Schematic atomic structures of the kink and antikink.
Green and blue David stars represent the left and right domains separated by the DW.
Red and yellow clusters represent unitcells of the domain wall translated by a kink (an antikink).
(d)~Recombination of the kink and antikink.
(e) and (f)~Detailed atomic configurations of a kink and an antikink. The conventional notation of Ta atoms in the CDW unitcell is shown in (f) with gray.
(g) and (h)~ Relaxed atomic structures of a kink and an antikink from the DFT calculations. Green and yellow balls represent Ta and S atoms. Black and magenta circles indicate Ta atoms associated with a kink or an antikink as in (c) and (d).
(i) and (j)~Simulated STM images of the kink and antikink. The images were obtained from the integrated charge density from -0.6 to 0~eV at a constant height of 2~$\AA$ away from the top S atom.
Green circles and ovals in (a), (b), (i), and (j) indicate the DW and DW-kink clusters, respectively.
}
  \label{stm}
\end{figure}

Figure~\ref{stm}(a) and \ref{stm}(b) show two different kinks, a kink and an antikink, one kinked in an opposite direction, shifting the DW4 in the transverse direction by one CDW unitcell (for the details, see Figure~S1 in Supporting Information). 
Due to the lack of mirror or $C_2$ symmetry, the antikink is distinguished structurally from the kink.
Longitudinally along the DW, the kink and the antikink shift the DW unitcell by a half unitcell, making the $+\pi$ and $-\pi$ phase shifts, respectively [red and yellow dashed lines in Figure~\ref{stm}(c)].
These phase shifts are one essential property of solitons. 
One can trace the atomistic details of the kinks based on the known atomic structure of the DW4 as shown in Figure~\ref{stm}.
	DW4 has a unitcell of 9 Ta atoms as marked with red or yellow colors in Figure~\ref{stm}(c).
	A kink has two distinct atomic configurations; two Ta atoms not belonging to the unitcell clusters of domains and DW [solid black circles in Figure~\ref{stm}(c) and white circles with number 6 and 11 in Figure~
 \ref{stm}(e)] and one Ta atom shared by a DW unitcell and a David-star unitcell of a domain edge [a dashed black circle in Figure~\ref{stm}(c) and a blue circle with number 2 in Figure~\ref{stm}(e)].
	In contrast, an antikink has three left-over Ta atoms [solid magenta circles in Figure~\ref{stm}(c) and white circles with number 2, 8, and 9 in Figure~\ref{stm}(f)] (for the STM images of a kink and an antikink overlaid with these atomic configurations, see  Figure~S1 in Supporting Information).
    These distinct atomic configurations are consistent with the fully optimized structures within the density functional theory (DFT) calculations [Figure~\ref{stm}(g) and \ref{stm}(h)], where the fine relaxations of local bonds are determined (for more details, see Supplemental Text 1 in Supporting Information).
	The STM topographies, reflecting the corrugation of the topmost sulphur layer, of the kinks are well reproduced by the simulated images based on these optimized structures [Figure~\ref{stm}(i) and \ref{stm}(j)].
    The sulphur atoms on the kink or the antikink form two distinct clusters ($\alpha$ and $\beta$, or $\alpha'$ and $\beta'$) as deformed from the pristine clusters of the CDW domain and the DW as indicated by green ellipses in both STM images measured [Figure~\ref{stm}(a) and ~\ref{stm}(b)] and simulated [Figure~\ref{stm}(i) and ~\ref{stm}(j)].
    That is, a kink and an antikink have different structures with different number of Ta atoms or electrons.

    To discuss the recombination of a kink and an antikink with an analogy of the pair ahnnihilation of Z$_2$ solitons, we consider a kink and an antikink with a separation of two CDW unitcells as shown in Figure~\ref{stm}(c).
    When the antikink is translated upward by two unitcells (or the kink is translated downward by two unitcells), the kink and the antikink combine into a perfect David-star unitcell and the DW4 is connected seamlessly [Figure~\ref{stm}(d)].
    This recombination verifies their relative phase shifts of $+\pi$ and $-\pi$, which indicates their solitonic or topological origin.
    Note but a clear difference from Z$_2$ solitons for only two degenerate ground states, where the soliton is its own antisoliton. 
    That is, two conventional Z$_2$ solitons will annihilate each other while geometrical kinks of the present system can be accumulated.
    This is due to to the well-known Z topology of geometrical kinks.

\subsection{Local electronic states of the kink}

\begin{figure}
  \includegraphics[width=17.3 cm]{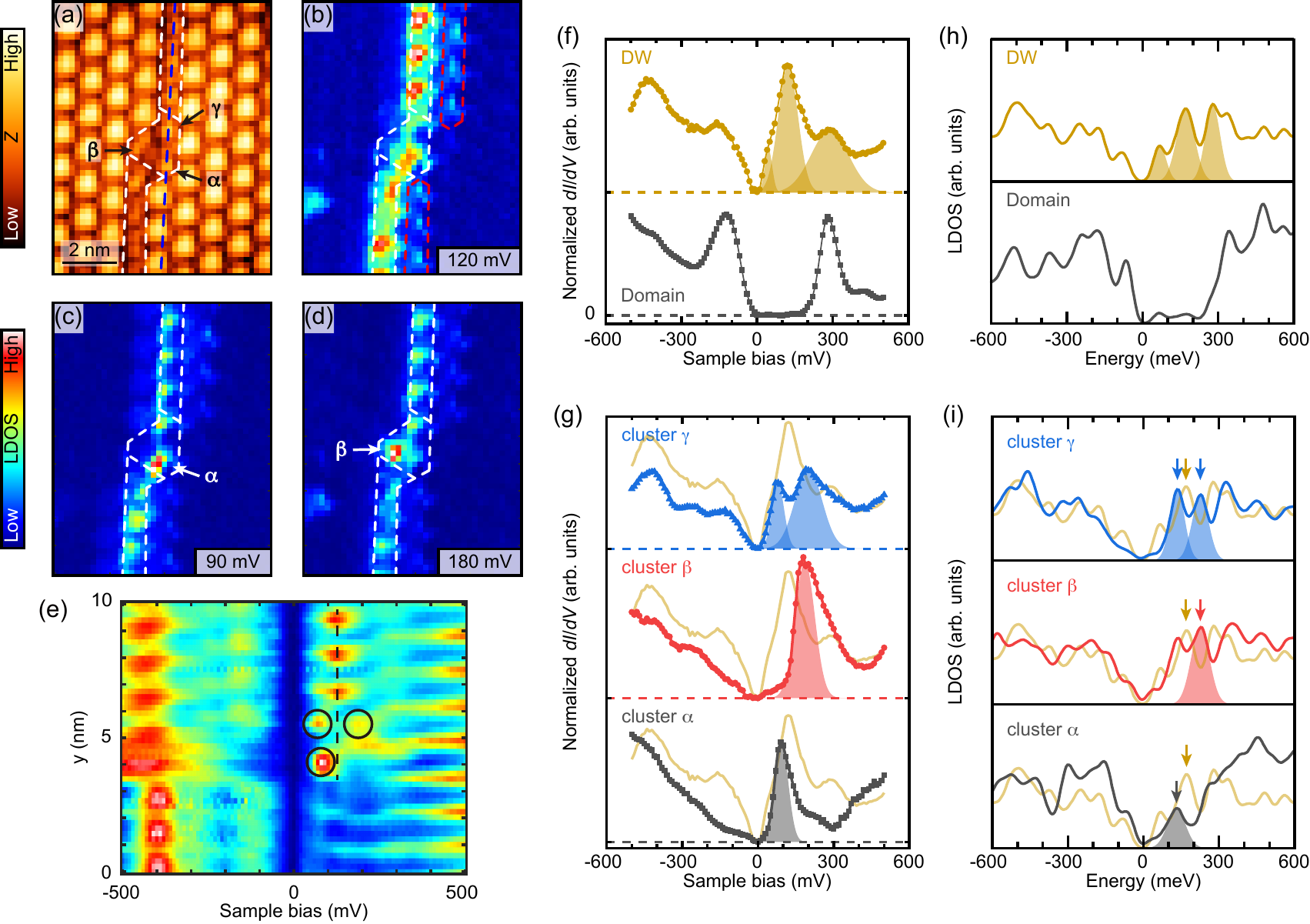}
  \caption{Local electronic states at the kink.
(a)~STM image a domain wall DW4 with a kink.
(b)-(d)~Corresponding LDOS maps at 120, 90, and 180~mV sliced from a full bias $dI/dV$ map. The DW and kink are guided with white dashed lines. Specific atomic clusters $\alpha$, $\beta$, and $\gamma$ within the kink are indicated by arrows.
(e)~Normalized $dI/dV$ spectra along the DW [dashed blue line in (a)]. A dashed line in (e) indicates the energy of the DW in-gap states and circles highlight the kink states of the clusters $\alpha$ and $\gamma$.
(f)~Normalized point $dI/dV$ spectra obtained at the CDW domain and DW.
(g)~Normalized $dI/dV$ spectra obtained at each cluster of the kink.
(h) and (i)~Corresponding LDOS spectra from the DFT calculations.
Note that in (g) and (i), we plot the duplicates of the DW spectra from (f) and (h) for comparison.
}
  \label{sts}
\end{figure}

The solitonic nature of DW kinks is further manifested by their electronic states.
The local density of states (LDOS) for the kink is directly probed by the differential tunneling conductance ($dI/dV$) with STM as shown in Figure~\ref{sts} (for the same measurements on the antikink, see Figure~S2 in Supporting Information).
  As clearly shown in the STM image and LDOS maps in Figure~\ref{sts}(a)-\ref{sts}(d), the kink clusters $\alpha$ and $\beta$ have distinct and well localized electronic states at around 90 and 180~meV, respectively, above the Fermi level. 
  These states are within the band gap of the CDW domains, which is located between about -100 and 300~meV [a black plot in Figure~\ref{sts}(f)], and are distinct from the in-gap states of the DW as seen in the 1D LDOS plot along the DW direction [Figure~\ref{sts}(e)]. 
  The DW has its in-gap states at 120~meV above the Fermi level [Figure~\ref{sts}(f)], and clearly noticed as an 1D electron channel in the LDOS map of Figure~\ref{sts}(b), which is consistent with our previous work~\cite{Cho2017}.
 In addition to the representative kink clusters $\alpha$ and $\beta$, the other part of the kink [the cluster $\gamma$ in Figure~\ref{sts}(a)] exhibits the states at 90 and 180~meV with similar intensities [Figure~\ref{sts}(g)].

    The main features of the kink electronic states are reasonably reproduced by DFT calculations.  
	In our calculations, the DW4 structure exhibits its major in-gap state at +170~meV [Figure~\ref{sts}(h)] (see also Figure~S3 in Supporting Information).
	The kink states with the optimized structure mentioned above appear at 137~meV for all clusters and 230~meV for the cluster $\beta$ and $\gamma$ [Figure~\ref{sts}(i)].
	Except for an overall energy difference of about 40 meV (due possibly to the extra charge transferred from the sublayer) and the suppression of the lower energy state for the cluster $\beta$, the consistency between the theory and the experiment is compelling if one consider the huge number of atoms involved in the calculation.
     Because all Ta \textit{d} electrons in the DW unitcells are paired and form the DW gap~\cite{Cho2017,ParkJH2021}, the kink, which is just the imperfect unitcell of the DW as discussed with Figure~\ref{stm}, has unpaired \textit{d} electrons. Thus, the kink in-gap states are manifested by these unpaired left-over \textit{d} electrons.

Combining all our STM and STS observations, we can verify the solitonic origin of the kink. There are a few criteria from the formal definition of a soliton as discussed in the previous works~\cite{Cheon2015,KimTH2017,ParkJH2022,KimTH2012}, such as (i) the existence of a translational phase shift in structure, (ii) the mobility of this phase defect, (iii) the existence of the band gap of the 1D chain, (iv) the localized in-gap states at the phase defect, and (v) the particle-antiparticle (soliton-antisoliton) annihilation. All these conditions are satisfied by the DW kink in the present work.

 \subsection{Topological origin of the kink}

The localized kink states can be understood by the emergence of edge states in the DW 1D electronic channel.
In order to examine the topological property of the DW and its edge modes, we construct a simplified tight binding model.
One complication here is that the electronic states of the DW have significant hybridization with one edge of the neighboring domain [red dashed lines in Figure~\ref{sts}(b)] (see also Figure~S3 in Supporting Information).
Therefore, we have to consider the DW clusters and their neighboring clusters (here we call domain edge clusters) together in the minimal model. 
The DW and domain edge clusters have alternating hopping amplitudes due to the different intercluster distances of 8.99 and 10.24~$\AA$, and 1D bands of the DW and domain edge are hybridized at M point (Figure~S3 in Supporting Information).
The model then becomes an asymmetric zigzag chain with alternative strong and weak hoppings [Figure~\ref{tb}(a)].
The hopping parameters are extracted by approximating the DFT band structure [Figure~\ref{tb}(b)].
This model yields four chiral edge modes: two edge terminations (the strong and weak hopping terminated) with two edge modes for each due to the double chain [Figure~S4 in Supporting Information].      
The kink is the combination of two edges with different terminations as shown in Figure~\ref{tb}(c) (for the details, see Figure~S4 in Supporting Information), as is its eigenstate spectrum [Figure~\ref{tb}(f)].
The weights and the energies of edge mode wave functions [Figure~\ref{tb}(d), \ref{tb}(e), and \ref{tb}(f)] are reasonably consistent with those of the kink states observed in our experiments [Figure~S7 in Supporting Information].
Namely, the in-gap state between the first and second lowest states [Figure~\ref{tb}(d)] is localized at the cluster $\alpha$ and $\gamma$ while that above the second lowest state [Figure~\ref{tb}(e)] is spread over three clusters with the largest weight at the cluster~$\beta$.
In spite of the oversimplified nature of the present model and the complex kink atomic structure excluded, we can learn that the kink states emerge from the topological edge modes of the 1D zigzag chain with the alternating hopping amplitudes.

However, we also find that the asymmetry of DW and domain edge chains (blue and red balls in the model) breaks the exact topology of the system, which is, otherwise, protected by the $C_2$ symmetry.
Nevertheless, the topological origin of the present kinks does not disappear.
Our calculation shows that the present in-gap states on the $\alpha$ and $\gamma$ clusters are directly connected to the topological edge mode when the $C_2$ symmetry is adiabatically recovered [Figure~\ref{tb}(g)] (for the details, see Supplemental Text 2 in Supporting Information).
Namely, the kinks inherit their stability and the in-gap states from the topological edge modes. 
At least, our kink states are well-separated from other bulk states in energy and hence robust against small perturbations~\cite{Benalcazar2017,ParkMJ2019,Qi2011}.
Other characteristics of a kink observed, such as the mobility and the soliton-antisoliton symmetry, are consistent with the topological picture.

\section{Conclusion}
We disclose electronic and topological properties of kinks along electronic domain walls in a van der Waals material of 1\textit{T}-TaS\textsubscript{2} with a correlated CDW phase.
Well-defined kinks and antikinks are identified along the CDW DWs by STM, which are mobile between pinning defects. 
Their atomic structures and in-gap electronic states are clearly resolved in STM and STS, which are well reproduced in our atomic structure models within the DFT calculations. 
Through the tight-binding approximation, these kinks and antikinds are mapped into Su-Schrieffer-Heeger topological solitons. 
Since various different domain walls can be created in a metastable way~\cite{Cho2017,ParkJH2021}, a variety of distinct kinks, at least 12 different DWs and 24 distinct kinks and antikinks, could in principle be formed (for the characterization of a kink along the other popular type of DWs, see Figure~S8 in Supporting Information).
Considering the pulse-induced DW creation~\cite{Cho2016}, controlling the creation and annihilation of kinks and antikinks seems to be promising, and it would be an important step toward the functionality of these kink states.
One fundamental characteristics of a kink is its \textit{Z} topology in geometry, distinct from the \textit{Z}\textsubscript{2} solitons, which makes the accumulation many kinks in a single DW possible as observed [Figure~\ref{mobile}(a)].
With a kink carrying a single unit of information, this property suggests that a kinked DW can have a cascade of such information units, which is an essential ingredient for multilevel computation such as synaptic devices~\cite{Seo2020}.
The robustness of DWs and kinks, the reversible processes involved, and their dissipationless kinetic nature have technological implications for, for example, a highly robust synaptic devices based on relatively robust and easily accessible van der Waals materials.


\section{Methods}
\threesubsection{STM and STS experiments}
1\textit{T}-TaS\textsubscript{2} single crystal was cleaved in high vacuum for a clean surface.
	The sample was pre-cooled at 78~K for 30~minutes and transferred to a cryogenic STM (4.4~K) kept in ultra-high vacuum for the STM~/~STS measurements.
	STM images were obtained by constant-current mode with a sample bias of -600~mV and the set current of 30~pA.
	We kept a small tunneling current (a large junction resistance of $\sim$10~G$\ohm$) to avoid the kink from moving while the scanning.
	Spectroscopic data were obtained using a lock-in amplifier with a frequency of 1~kHz and a modulation of 5~mV.\\

\threesubsection{DFT calculations}
DFT calculations were performed by using the Vienna \textit{ab initio} simulation package~\cite{Kresse1996} within Perdew-Burke-Ernzerhof generalized gradient approximation~\cite{Perdew1996} and the projector augmented wave method~\cite{Blochl1994}.
The single-layer 1\textit{T}-TaS\textsubscript{2} was modeled with a vacuum spacing of about 24~$\AA$.
We used a plane-wave basis set of 259~eV and a $6\times6\times1$ \textit{k}-point mesh for the $\sqrt{13}\times\sqrt{13}$ CDW unitcell.
All atoms at the kink were relaxed until the residual force components were within 0.02~eV/$\AA$ and the atoms at domain and DW are fixed at their own position to avoid unexpected relaxation effect due to the periodic supercell approximation.
To more accurately represent electronic correlations, the on-site Coulomb energy ($U$~=~2.3~eV) was included for Ta~5$d$ orbitals.\\

\begin{figure}
  \includegraphics[width=17.7 cm]{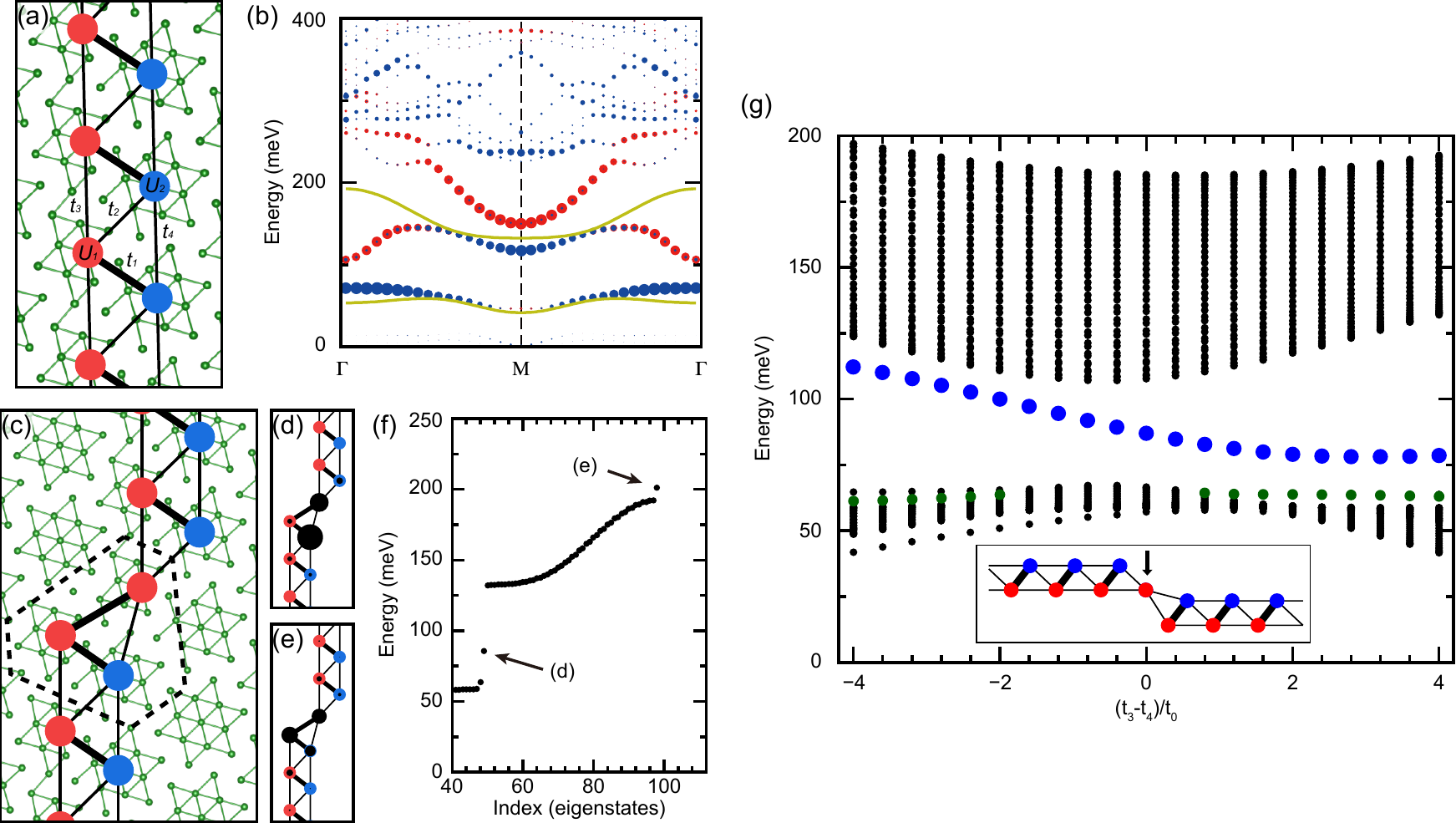}
  \caption{In-gap states of the kink.
(a)~Schematic image of the tight-binding model of the DW4.
(b)~Comparison between the DFT (red and blue dots) and tight-binding (yellow line) band structure, where $t_0$=0.009, $t_1$=4.55$t_0$, $t_2$=2.36$t_0$, $t_3$=3$t_0$, $t_4$=-$t_0$, $U_1$=0.10, and $U_2$=0.11.
(c)~A schematics for the tight-binding model for the kink.
We mark the clusters at the kink with a dashed hexagon. 
Note that the hopping amplitude between the cluster $\beta$ and $\gamma$ is enhanced by a factor of 2 (2$t_3$) for better comparison (see Supplemental Text 3, Figure~S5, and Figure~S6 in Supporting Information).
(d) and (e)~Wave functions for the in-gap states. Each eigenstate is marked in (f).
(f)~Energy spectrum calculated in this model.
The lowest state is highly localized at the cluster $\alpha$ and $\gamma$ while the second-lowest state is mainly localized at the cluster $\beta$ with non-negligible weights at the cluster $\alpha$ and $\gamma$.
(g)~Adiabatic evolution of topological edge state for the asymmetric kink chain.
Eigenstate energies (black circles) are plotted as a function of the difference between hopping amplitudes of two asymmetric chains [($t_3$-$t_4$)/$t_0$] and total hopping amplitude of $t_3$+$t_4$ is fixed at 2$t_0$.
($t_3$-$t_4$)/$t_0$=0 corresponds to the symmetric point [$t_3$=t$_4$=$t_0$ (0.009)] and ($t_3$-$t_4$)/$t_0$=4 corresponds to the kink structure in the DW4 ($t_3$=3$t_0$, $t_4$=-$t_0$).
Blue and green circles denotes the adiabatically connected edge state of the kink and perturbed states localized at around kink, respectively. 
Inset is a schematic model of the kink and the arrow denotes the $C_2$ rotation center.
}
  \label{tb}
\end{figure}

\threesubsection{Tight-binding calculations}
The PythTb package was used for the tight-binding calculations~\cite{PythTb}.
We constructed a tight-binding model for the 1D zigzag chain at the DW. The tight-binding Hamiltonian of the ziazag chain with two sublattices within each unit cell is 
\begin{equation}
    H = \sum_{i}\left (  t_{1}c_{Ai+1}^{\dagger}c_{Bi} + t_{2}c_{Ai}^{\dagger}c_{Bi} + t_{3}c_{Ai}^{\dagger}c_{Ai+1} + t_{4}c_{Bi}^{\dagger}c_{Bi+1} + h.c. \right ) + \sum_{i}\left (  U_{1}c_{Ai}^{\dagger}c_{Ai} + U_{2}c_{Bi}^{\dagger}c_{Bi} \right ) \label{TB},
\end{equation}
where $c_{i}^{\dagger}$ and $c_{i}$ are creation and annihilation operators, and A and B represent two sites of a unit cell [Figure~\ref{tb}(a)]. The parameters of the hopping amplitudes $t_{i}$ and the on-site energies of $U_{i}$ are adjusted to fit the DFT band structure without the extra electron correlation [Figure~\ref{tb}(b)] because the DW states are not significantly changed with the extra electron correlation (Figure~S3 in Supporting Information). The alternating hopping amplitudes of $t_{1}$ and $t_{2}$ are reflected in the intercluster distances of 8.99 and 10.24 {\AA} and transition metal scale as $d^{-5}$, and we used the asymmetric the hopping parameters along the chain direction ($t_{3}$ and $t_{4}$) to describe the broken DS cluster at DW and perfect DS cluster at domain boundary.
We obtained the hopping and on-site parameters of $t$=0.009, $t_1$=4.55$t$, $t_2$=2.36$t$, $t_3$=3$t$, $t_4$=-$t$, and $U_1$=0.10, $U_2$=0.11.



\medskip
\textbf{Supporting Information} \par 
Supporting Information is available from the Wiley Online Library or from the author.

\medskip
\textbf{Acknowledgements} \par 
This work was supported by the Institute for Basic Science (Grant No. IBS-R014-D1).

\medskip
\textbf{Conflict of Interest} \par 
The authors declare no conflict of interest.

\medskip

%
\bibliographystyle{MSP}



\clearpage


\end{document}